\documentstyle[11pt,epsfig]{article}
\def\be{\begin{eqnarray}}
\def\ee{\end{eqnarray}}

\def\J#1#2#3#4{ {#1} {\bf #2}, {#3} (#4) }
\def\PRL{Phys. Rev. Lett.}

\def\PLB{Phys. Lett. B}

\def\NPA{Nucl. Phys. A}
\def\NPB{Nucl. Phys. B}

\def\PRC{Phys. Rev. C}
\def\PRPTS{Physics Reports}
\def\bea{\be}
\def\eea{\ee}

\def\roughly#1{\mathrel{\raise.3ex\hbox{$#1$\kern-.75em%
\lower1ex\hbox{$\sim$}}}}

\def\gsim{\roughly>}

\def\PRL{Phys. Rev. Lett.}

\def\PLB{Phys. Lett. B}

\def\NPA{Nucl. Phys. A}
\def\NPB{Nucl. Phys. B}

\def\PRC{Phys. Rev. C}

\begin{document}

\renewcommand{\thefootnote}{\fnsymbol{footnote}}
\setcounter{footnote}{0}

\vskip 0.4cm
\hfill {\bf KIAS-P99044}

\hfill {\today}
\vskip 1cm

\begin{center}
{\LARGE\bf Effective Field Theory For Nuclei,}
\vskip 0.2cm

{\LARGE\bf Dense Matter And The Cheshire Cat}\footnote{Talk given at the 12th
Nuclear Physics Summer School and Symposium and the 11th International
Light-Cone Workshop ``New Directions in QCD," 21-25 June 1999, Kyungju,
Korea}

\date{\today}

\vskip 1.5cm

{\large
Mannque Rho
}

\end{center}

\vskip 0.5cm

\begin{center}

{\it School of Physics, Korea Institute for Advanced Study,
Seoul 130-012, Korea}

{and}

{\it Service de Physique Th\'eorique, CE Saclay,
91191 Gif-sur-Yvette, France}\footnote{Permanent address}

{E-mail: rho@spht.saclay.cea.fr}

\end{center}

\vskip 0.5cm

\begin{abstract}

In this talk I discuss three related topics based on some of the recent
developments in hadron and nuclear physics: one, effective field
theory approach to two-nucleon systems; two, an explanation of the
flavor singlet axial charge in the proton (i.e., ``proton spin
problem") in terms of a Cheshire Cat phenomenon; and three, the
quark-hadron duality in hadronic matter at high density and
``qualitons" at high density (``superqualitons''). The principal
common theme in these discussions will be the emergence of the
generic feature of the Cheshire Cat Principle.

\end{abstract}

\newpage

\renewcommand{\thefootnote}{\#\arabic{footnote}}
\setcounter{footnote}{0}

\section{Introduction}\label{intro}
\indent\indent
I would like to discuss in this talk three topics which
superficially would look unrelated but in essence can be connected
by the general theme of Cheshire Cat. I will first look at
two-nucleon systems in terms of macroscopic variables of QCD,
namely hadrons; I will then go back to an elementary hadron, in
particular the proton and examine its microscopic structure by
``punching" a hole in the proton and then putting quarks and gluons
of the appropriate quantum numbers inside and argue that there is a
continuity in the descriptions in terms of hadronic variables and
in terms of quark-gluon variables; finally the continuity of quarks
and hadrons at high density will be described in terms of qualitons
-- quark solitons. Many of my collaborators have contributed to the
development discussed in this talk. Among them are Gerry Brown,
Bengt Friman, Deog Ki Hong, Kuniharu Kubodera, Kurt Langfeld,
Hee-Jung Lee, Dong-Pil Min, Byung-Yoon Park, Tae-Sun Park, Vicente
Vento and Ismail Zahed.
\section{Two-Nucleon Systems in Effective Field Theory}
\indent\indent
Consider two nucleon interactions at very low energy. We are
generically interested in the energy-momentum probe much less than
the pion mass $m_\pi\sim 140$ MeV. At this energy, according to
Weinberg's ``theorem," the content of QCD can be phrased in terms
of the nucleons and pions. In fact if we are probing a scale much
less than the pion mass, we can even ignore the pions and work with
the nucleons only. The corresponding framework is an effective
field theory (EFT). Much work has been done on this EFT for
two-nucleon systems~\cite{weinberg,pmr,vankolck,KSW,PKMR} (see
\cite{EFTmeetings} for recent reviews). There are two classes of
observables to look at. One is scattering process and the other is
response function to external fields. The two are of course
complementary in revealing the physics involved. Recent efforts
have been put more on scattering than on response functions
although more information has traditionally been gained from the
latter in nuclear physics. Both need to be treated simultaneously
which I will do here.

In the literature so far, there are broadly two approaches to EFT
in nuclear physics. One is the original Weinberg
approach~\cite{weinberg} where a systematic power counting is made
only to the ``irreducible graphs,'' for which chiral perturbation
theory (with pions figuring crucially) becomes applicable in
organizing the expansion and the reducible graphs are summed to all
orders with the irreducible graphs entering as vertices. This
scheme used in Ref.\cite{weinberg,pmr,vankolck,PKMR} -- which in
spirit is close to the original Wilsonian EFT but incurs possible
errors in the power counting -- involves a scale $\Lambda\gsim
m_\pi$ as the counting is applied only to the irreducible terms. I
will call this the $\Lambda$ counting. The other
approach~\cite{KSW} motivated to account more transparently for the
large s-wave scattering lengths in two-nucleon scattering purports
to do a systematic counting for the S-matrix as a whole which
amounts to summing all graphs involving leading non-derivative
four-Fermi contact interactions while treating all others,
including pion exchanges, as perturbation. This approach renders a
more systematic accounting of the powers of $Q/\Lambda$ where
$Q=\sqrt{MB}$, $p$ as well as $m_\pi$ but at the expense of certain
predictivity. This is referred to as $Q$-counting scheme.

There have been lots of hot debates as to whether one scheme is
more powerful and or more consistent than the other, mainly in
connection with the scattering~\cite{EFTmeetings}. While the
situation is not completely settled, I believe that it is fair to
say that the two are both consistent with the tenet of EFT and
roughly equivalent in its power. Eschewing the debate which seems
somewhat academic, I will simply focus on the $\Lambda$-counting
approach in this talk. In fact, in the processes that I will
consider, I would claim that the $\Lambda$ counting is more
adaptable to --- and more predictive in treating
-- nuclear physics problems than the $Q$ counting. One great
advantage of the $\Lambda$ scheme is that it allows one to
calculate precisely defined corrections to what can be obtained
from so-called realistic potential models (PM in short) that have
been developed by nuclear theorists since a long time, thus giving
the realistic potential models (PM) a first-principle
justification. It allows us to study processes involving not just
few-body but also many-body systems. For instance, it is possible
to calculate the ``hep" process in the Sun $p+^3{\rm He}\rightarrow
^4{\rm He}+e^++\nu_e$ (which is currently an
exciting issue after the recent Surperkamiokande neutrino data)
with an accuracy that can be controlled systematically. It has also
been successfully applied to calculating axial charge transitions
of heavy nuclei (see \cite{rho}) with the additional ingredient of
BR scaling~\cite{BR}.

While there is no definitive evidence that the $\Lambda$ scheme is
fully justified for n-body systems with $n\gg 2$, it definitely
works for $n=2$ systems. In Ref.\cite{PKMR}, it has been shown in a
cutoff regularization (with a finite cutoff as required by
EFT~\cite{lepage}) that the EFT results of the leading order terms
in all two-body observables at low energy $E\ll m_\pi$ are
precisely reproduced by the potential model results. This is the
case not only for scattering amplitudes but also all electroweak
response functions. What EFT brings in addition to what we get from
the PM is a systematic procedure to compute corrections to the
leading order results. For low-energy processes, this status of the
PM can be understood by the fact that the tail of the wave
functions is a physical quantity and the realistic potential models
which are fit to experiments have the {\it correct} asymptotic
properties in the wave function. This point has also been stressed
and clarified by Phillips and Cohen in a recent important paper~\cite{pcohen}.

Considered to order $Q^n$ where $n$ is the order in the $\Lambda$
counting (which I will consider relative to the leading order term
in the expansion of the irreducible graphs), the s-wave scattering
amplitudes are accurately postdicted~\cite{PKMR,hmp} up to $p\leq
m_\pi$ for $n=2$ and a cutoff appropriate to the number of pions
exchanged (one or two) in the irreducible graphs. Deuteron
properties are also well understood within the same
scheme~\cite{PKMR}.  The scheme allowed the calculations to order
$Q^2$ and $Q^3$ of the proton fusion process in the Sun~\cite{pp}
\bea
p+p\rightarrow d+e^+ +\nu_e\label{pp}
\eea
and of the threshold np capture~\cite{pmr,pkmr99} with polarized
projectile and target nucleons
\bea
\vec{n}+\vec{p}\rightarrow d+\gamma.\label{np}
\eea
The process (\ref{pp}) crucial for the solar neutrino problem is
given in the scheme to an accuracy of $1\sim 3$ percents (the
uncertainty here is due to the exchange current that appears at
order $Q^3$). The unpolarized cross section for (\ref{np}) has been
computed to the accuracy of 1 percent in a complete agreement with
the experiment. More significantly, the polarization observables
$P$ (circular polarization) and $\eta$ (anisotropy) have been
predicted {\it parameter-free}
in Ref.\cite{pkmr99}\footnote{A similar prediction in the
$Q$ scheme was made by Chen, Rupak and Savage~\cite{CRS}.}. This is
a genuine prediction since there are no experimental data
available. (They are currently being measured in ILL of
Grenoble~\cite{ILL}.)

In all these postdictions and predictions, there is very little
$\Lambda$ dependence as required by the tenet of EFT. This is a
clear indication that the scheme is fully consistent.

 One can go up in the momentum range by doing higher
order calculations. Phillips and Cohen~\cite{pcohen} discuss how
the two-body EM form factors can be described in the $\Lambda$
scheme. Pushing somewhat the validity of the scheme, one can
calculate even the process
\bea
e+d\rightarrow e+n+p
\eea
involving large momentum transfers $q\gsim 1$ GeV. In fact this
process measured in 1980's at ALS of Saclay and elsewhere is
considered to be the {\it unambiguous} confirmation of
meson-exchange currents in nuclei (see
~\cite{frois}).\footnote{When I suggested this process at the
second INT-Caltech EFT meeting as a case for testing the EFT
strategy in the $Q$ counting, everyone (!) in the audience chuckled
and said the process is completely out of reach for EFT (see
\cite{seki}). I grant that this may be true in the $Q$ scheme at
least for the moment but {\it not} in the $\Lambda$ scheme where it
has worked stunningly, confirming the ``chiral filter hypothesis."
{\it Voil\`a} the power of the $\Lambda$ scheme!}
\section{``Proton Spin" and The Cheshire Cat}
\indent\indent
The nucleons that figured in the above section were color-singlet
point-like fields that say nothing explicit about quarks and
gluons. Let me now imagine puncturing a hole of size of radius $R$
in the proton and populating the inside with the QCD degrees of
freedom, quarks and gluons. How to do this consistently with QCD is
known and the model involved is the chiral bag (this is described
extensively in ~\cite{NRZ}) which consists of a quark-gluon sector
inside the bag and a color-singlet hadronic sector outside, with
the two sectors connected by suitable boundary conditions. When
constructed with relevant degrees of freedom and in consistency
with the symmetries of QCD, the model gives what is now called
``Cheshire Cat." In short, the Cheshire Cat Principle~\cite{NRZ}
states that at low energy, physics involving hadrons should be
independent of the bag size $R$. It has been shown that this
principle is operative semi-quantitatively in {\it all} properties
of the nucleon~\cite{toki} {\it except for} the flavor-singlet
axial charge (FSAC) of the proton which is related to the ``proton
spin." Here I will briefly describe -- leaving the details to the
paper by Lee et al~\cite{heejung} --  how this Cheshire Cat property
can be recovered in the FSAC when chiral symmetry and chiral
anomaly are judiciously taken into account. It turns out that the
interplay between the boundary conditions and Casimir effects plays
a crucial role.

\begin{figure}
\centerline{\epsfig{file=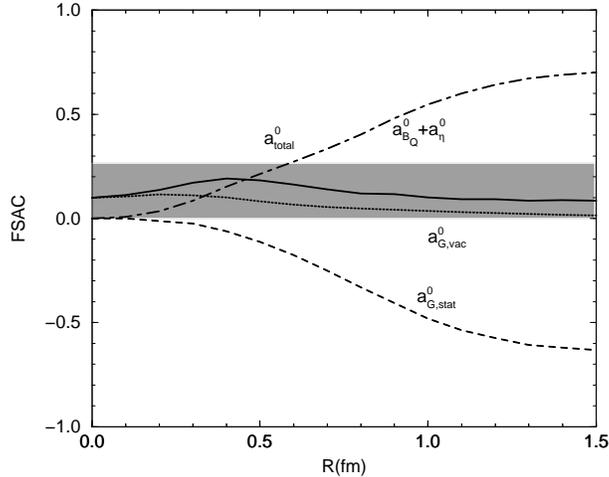, width=9cm}}
\caption{The flavor singlet axial charge
of the proton as a function of the bag radius compared with the
experiment; it consists of three contributions: (a) {\it matter
field contribution}: quark plus $\eta$ ($a^0_{B_Q} + a^0_\eta$),
(b) {\it gauge field contribution}: the static gluons coupled to
the quark source ($a^0_{G,stat}$), (c) {\it Casimir contribution}:
the gluon vacuum fluctuation ($a^0_{G,vac}$), and (d) the sum total
($a^0_{total})$. The shaded area represents the range admitted by
experiments.}\label{fsac}
\end{figure}

Since the flavor-singlet axial current is not conserved because of
the anomaly, the color cannot be confined inside the bag unless a
suitable boundary condition is put at the
surface that cancels the outflow of the color~\cite{NRWZ}. The
boundary term that does this is proportional to the Chern-Simons
current on the surface, i.e., the  Chern-Simons flux (which is invariant
under neither small nor large gauge transformation). This influences the
FSAC of the proton nontrivially. Briefly, what happens is that the
FSAC contributed by the matter fields (quarks inside the bag and
$\eta^\prime$ outside the bag) and the FSAC contributed by the
gauge field (gluons inside the bag) more or less (or possibly
exactly if treated rigorously) cancel, leaving behind only the
small contribution from the (gauge field) vacuum fluctuation which
is effectively a Casimir effect caused by the boundary with its
color-anomalous boundary condition. The cancellation and the
remnant small FSAC are shown in Fig. \ref{fsac}.

The upshot of the small result left over which is independent of 
the size $R$
provides yet another compelling evidence that the proton could
be equivalently understood {\it both} in terms of quarks/gluons
and in terms of macroscopic hadronic variables. When $R$ is taken to
be big, it is the QCD variables that take over, e.g., the MIT bag.
When the size $R$ is shrunk to a point, the proton is a skyrmion.
Thus we have the equivalence of the skyrmionic proton and the
quark-gluonic (QCD) proton, that is, the Cheshire Cat. Since there
is no way one can exactly bosonize four-dimensional QCD, the
equivalence cannot be made exact. We simply have an
approximate equivalence which can be made more precise by doing
more work. What determines the language to use is the kinematic
condition of the process one looks at.
\section{Dense Matter and The Cheshire Cat}
\indent\indent
The last topic I would like to discuss is infinite nuclear matter
at large density. The old lore that at an asymptotic density, the
matter can be described by perturbative QCD with weakly interacting
boring quarks is simply wrong. What may be happening at high
density is something a lot more exciting than that. This explains
why lots of people are presently writing papers about it.

What is most surprising and in some sense unexpected is that at
high density the Cheshire Cat picture becomes more prominent. In
fact, at high density, there ceases to be any real distinction
between quarks and hadrons. This can be best seen in terms of a
quark soliton analogous to the qualiton Kaplan~\cite{kaplan}
introduced as a model for the constituent quark. The mechanism I
will discuss exploits that at high density diquarks condense giving
rise to color superconductivity as recently discussed in the
literature~\cite{CSC}. Since the resulting qualiton is formed from
a color superconducting ground state, it seems proper to call it
superqualiton~\cite{superqualiton}. It has been argued on general
symmetry and dynamical grounds~\cite{duality} that at high density,
hadronic matter of flavor $SU(3)$ is characterized by the
condensate
\be
\Big< q_{L\alpha}^{ia}q_{L\beta}^{jb} \Big>
=-\Big<q_{R\alpha}^{ia}q_{R\beta}^{jb}\Big>
=\kappa \,\,\epsilon^{ij}\epsilon^{abI}\epsilon_{\alpha\beta I}
\label{1}
\ee
where $\kappa$ is some constant, $i,j$ are $SL(2,C)$ indices, $a,b$
are color indices, and $\alpha,\beta$ are flavor indices. Equation
(\ref{1}) holds for parity-even states. Such a condensate locks
color and flavor so that global color and chiral symmetry are
broken to the diagonal subgroup $SU(3)_{C+L+R}$. The consequence of
this is that there is an invariant $U(1)$ subgroup that contains a
``twisted" photon, measured with which all excitations carry
integer charges reminiscent of the Han-Nambu quarks and have
quantum numbers that correspond to those of the mesons and baryons
present at zero density. There is then a continuity between the
excitations at high density in terms of quarks and gluons and
hadronic excitations at low density in terms of baryons and mesons.
This clearly is a case of Cheshire Cat.

Now to see that this is the Cheshire Cat in the sense formulated in
terms of the chiral bag~\cite{NRZ}, consider the excitation of a
quark on top of the diquark-condensed ``vacuum." In \cite{duality},
such a quark is argued to behave like a baryon. Now I claim that
this quark is a quark soliton, i.e.,
superqualiton~\cite{superqualiton}.

To describe the low-energy dynamics of the color-flavor locking
phase, introduce a field $U_L(x)$ which maps space-time to the
coset space, $M_L=SU(3)_c\times SU(3)_L/SU(3)_{c+L}$. One can take
it to be
\begin{equation}
{U_L}_{a\alpha}(x)\equiv\lim_{y\to x}{\left|x-y\right|^{\gamma_m}
\over\kappa}\epsilon^{ij}\epsilon_{abc}\epsilon_{\alpha\beta\gamma}
q^{b\beta}_{Li}(-\vec v_F,x)q^{c\gamma}_{Lj}(\vec v_F,y),\label{U}
\end{equation}
where $\gamma_m$ is the anomalous dimension of the diquark field of
order $\alpha_s$ and $q(\vec v_F,x)$ denotes the quark field with
momentum close to a Fermi momentum $\mu\vec v_F$. The pairing
involves quarks near the opposite edge of the Fermi surface.
Similarly, we introduce a right-handed field $U_R(x)$, also a map
from space-time to $M_R=SU(3)_c\times SU(3)_R/SU(3)_{c+R}$, to
describe the excitations of the right-handed diquark condensate. If
this field takes a vacuum expectation value as a consequence of the
diquark condensation which will, owing to (\ref{1}), have the form
\bea
\left<{U_L}_{a\alpha}\right>=-\left<{U_R}_{a\alpha}\right>=\kappa\,
\delta_{a\alpha},
\eea
then 16 Nambu-Goldstone bosons will get excited~\footnote{Actually 
there are 17 of them, one of which has to do with spontaneous breaking
of the baryon number.}. Eight of them will
get eaten up by the gluons to give masses to the gluons. The
massive gluons then turn into massive vector mesons whose quantum
numbers are those of the light-quark vector mesons present at zero
density. The remaining eight (pseudoscalar) Nambu-Goldstone bosons
are the equivalents of the ones present at zero density and are
represented by the interpolating field (\ref{U}). In analogy to the
usual skyrmion at zero density, this field supports a soliton which
is a fermion, the quantum numbers of which are identical to those
of the usual baryon.

The effective Lagrangian that gives rise to this soliton should in
principle be derived from QCD with the help of renormalization
group flows toward the Fermi surface of high quark density. Such a
derivation would determine the parameters that figure in the
effective Lagrangian such as the ``pion decay constant" $F$ etc
which would carry information on the superconductivity gap etc. At
the moment such an effective Lagrangian is not known, so a detailed
study of it excitation structure cannot be discussed.

However one can venture to make a few interesting conjectures.
Viewed as a superqualiton whose mass is given by the soliton mass,
there is nothing that requires that the soliton mass be equal to or
near the superconductivity gap $\Delta$ (which is dictated by the
condensate). In fact there is nothing which would prevent the mass
from being much less than the gap. Thus one could imagine that light
fermions are excited {\it within} the gap. Correlations between
light superqualitons could rearrange the ground state into a
different form from that of the standard superconductivity. For
this reason the phenomenon of color superconductivity in QCD at
high density could be completely different from the usual BCS
superconductivity. A similar point in a different context was
raised in \cite{pisarski}.
\section{Conclusion}
\indent\indent
The most important outcome of the recent development of EFT in
nuclear physics is that the highly successful approach to nuclear
structure using realistic nuclear potentials (PM) is rendered a
first-principle interpretation in that it represents the leading
term in the EFT expansion with the corrections thereof
systematically calculable. This confers the power of modern field
theory techniques to the standard nuclear physics approach that has been
practiced with success since a long time. I am suggesting that this
``bridging" comes about thanks to a possible duality between QCD
variables and macroscopic (color-singlet) variables that I refer to
as the Cheshire Cat Principle. The ``proton spin problem" is an
illustration of this in the basic structure of the hadron.

In the case of high density, the picture becomes even more
intriguing. There we see emerging the symbolic (approximate)
equality
\bea
``{\rm Quark}"\approx ``{\rm Qualiton}"\approx ``{\rm Baryon}".
\eea

 It is amusing that the notion of the Cheshire Cat which
was conceived by the need to reconcile the traditional
meson-exchange description with the modern QCD description for
nuclear processes~\cite{littlebag} (i.e., the ``little bag" with
pion cloud, chiral bag etc) at low density re-emerges at high
density where one would have expected the bona-fide QCD to be
uniquely applicable.

\subsection*{Acknowledgments}
\indent\indent
I would like to thank the organizers of this meeting, particularly
Chueng-Ryong Ji and Dong-Pil Min for inviting me to give this talk.
I would also like to acknowledge the hospitality of KIAS and its
Director, C.W. Kim while this paper was being written.

\end{document}